\documentclass[aps,prd,secnumarabic,nobibnotes,twocolumn,superscriptaddress]{revtex4-1}
\usepackage{amsfonts}
\usepackage{mathrsfs}
\usepackage{amsmath}
\usepackage{color}
\usepackage{natbib}
\usepackage{graphicx}
\usepackage{bm}
\usepackage{amssymb}
\usepackage{xspace}
\usepackage{epstopdf}
\usepackage{dcolumn}
\usepackage{multirow}
\usepackage[colorlinks=true, letterpaper=true, pdfstartview=FitV, linkcolor=blue, citecolor=blue, urlcolor=blue]{hyperref}
\usepackage{wrapfig}

\makeatletter

\newcommand{\Rmnum}[1]{\expandafter\@slowromancap\romannumeral #1@}
\makeatother

\begin{document}
\title{Coexistence of hourglass Weyl and Dirac nodal line phonons in orthorhombic-type KCuS}

\author{Jianhua Wang}\thanks{J.W. and H.Y. contributed equally to this manuscript.}
\address{School of Physical Science and Technology, Southwest University, Chongqing 400715, China.}

\author{Hongkuan Yuan}\thanks{J.W. and H.Y. contributed equally to this manuscript.}
\affiliation{School of Physical Science and Technology, Southwest University, Chongqing 400715, China.}

\author{Ying Liu}\thanks{Corresponding authors}\email{ying$_$liu@hebut.edu.cn}
\address{School of Materials Science and Engineering, Hebei University of Technology, Tianjin 300130, China}

\author{Feng Zhou}
\address{School of Physical Science and Technology, Southwest University, Chongqing 400715, China.}

\author{ Juntao Ma}\address{School of Physical Science and Technology, Southwest University, Chongqing 400715, China.}

\author{Minquan Kuang}\address{School of Physical Science and Technology, Southwest University, Chongqing 400715, China.}

\author{Tie Yang}\address{School of Physical Science and Technology, Southwest University, Chongqing 400715, China.}

\author{Xiaotian Wang}\thanks{Corresponding authors}\email{xiaotianwang@swu.edu.cn}
\address{School of Physical Science and Technology, Southwest University, Chongqing 400715, China.}

\author{Gang Zhang}\thanks{Corresponding authors}\email{zhangg@ihpc.a-star.edu.sg}
\address{Institute of High Performance Computing, Agency for Science, Technology and Research (A*STAR), 138632, Singapore}

\begin{abstract}
In parallel to electronic systems, the concept of topology has been extended to phonons, which has led to the birth of topological phonons. Very recently, nodal point phonons, nodal line phonons, and nodal surface phonons have been proposed, and a handful of candidate materials have been predicted in solid-state materials. In this work, based on symmetry analysis and first-principles calculations, we propose that hourglass Weyl nodal line (HWNL) phonons and Dirac nodal line (DNL) phonons coexist in the phonon dispersion of a single material, KCuS, with a $Pnma$-type structure. The HWNLs and DNLs are relatively flat in frequency and well separated from other phonon bands. The corresponding topological phonon surface states appear only in the [100] and [001] surfaces, unlike those of typical phononic nodal-line materials. The reason for this phenomenon is explained based on Zak phase calculations. Our work, for the first time, proves that phononic nodal lines with different types of degeneracies can be achieved in one single material. Thus, KCuS can be viewed as a good platform to investigate the entanglement between HWNL phonons and DNL phonons.
\end{abstract}
\maketitle

\section{Introduction}
Representing a new chapter in condensed matter physics and material science, topological quantum states~\cite{add1,add2} have attracted the wide attention of researchers since their discovery. The topological concepts of electronic structures have been widely predicted and verified ~\cite{add3,add4,add5,add6,add7,add8,add9,add10,add11,add12,add13,add14}. To date, thousands of topological electronic materials have been widely predicted and experimentally observed ~\cite{add15,add16,add17,add18,add19,add20,add21,add22,add23,add24,add25}. Moreover, phonons, the most basic emergent boson of crystalline lattices, are the energy quantum of lattice vibration. Phonons make important contributions to the thermal conductivity, and specific heat of non-metal. The coupling between phonons and electrons determines the properties of many physical phenomena, including superconductivity, thermoelectric effect and carrier mobility~\cite{add26,add27}. By analogy with well-studied electronic systems, topological concepts have been introduced to the field of phonons and named topological phonons~\cite{add28,add29,add30,add31,add32,add33}. It is worth mentioning that unlike electrons, phonons are bosons and are not restricted by the Pauli exclusion principle, so we can focus on the topological signatures of a wide frequency range.

Very recently, topological phonons in solid-state materials have been studied theoretically and experimentally. A series of three-dimensional (3D) materials with rich topological phononic states have been proposed: (i) different types of nodal point phonons~\cite{add34,add35,add36,add37,add38,add39,add40,add41,add42,add43,add44,add45}, including the single and high degenerate Weyl point phonons, ideal type II Weyl point phonons, unconventional triangular Weyl point phonons, Dirac point phonons, triply nodal point phonons, and six-fold nodal point phonons; (ii) different types of nodal line phonons ~\cite{add46,add47,add48,add49,add50,add51,add52,add53,add54}, including nodal-ring phonons, Weyl nodal straight line phonons, helical nodal line phonons, Weyl open nodal line phonons, and hourglass nodal-net phonons; and (iii) multiple nodal surface phonons ~\cite{add55}. Among them, double Weyl point phonons in parity-breaking FeSi ~\cite{add34} and phononic helical nodal lines in MoB$_2$ ~\cite{add48} have been verified via inelastic X-ray scattering.

\begin{figure}
\includegraphics[width=7.8cm]{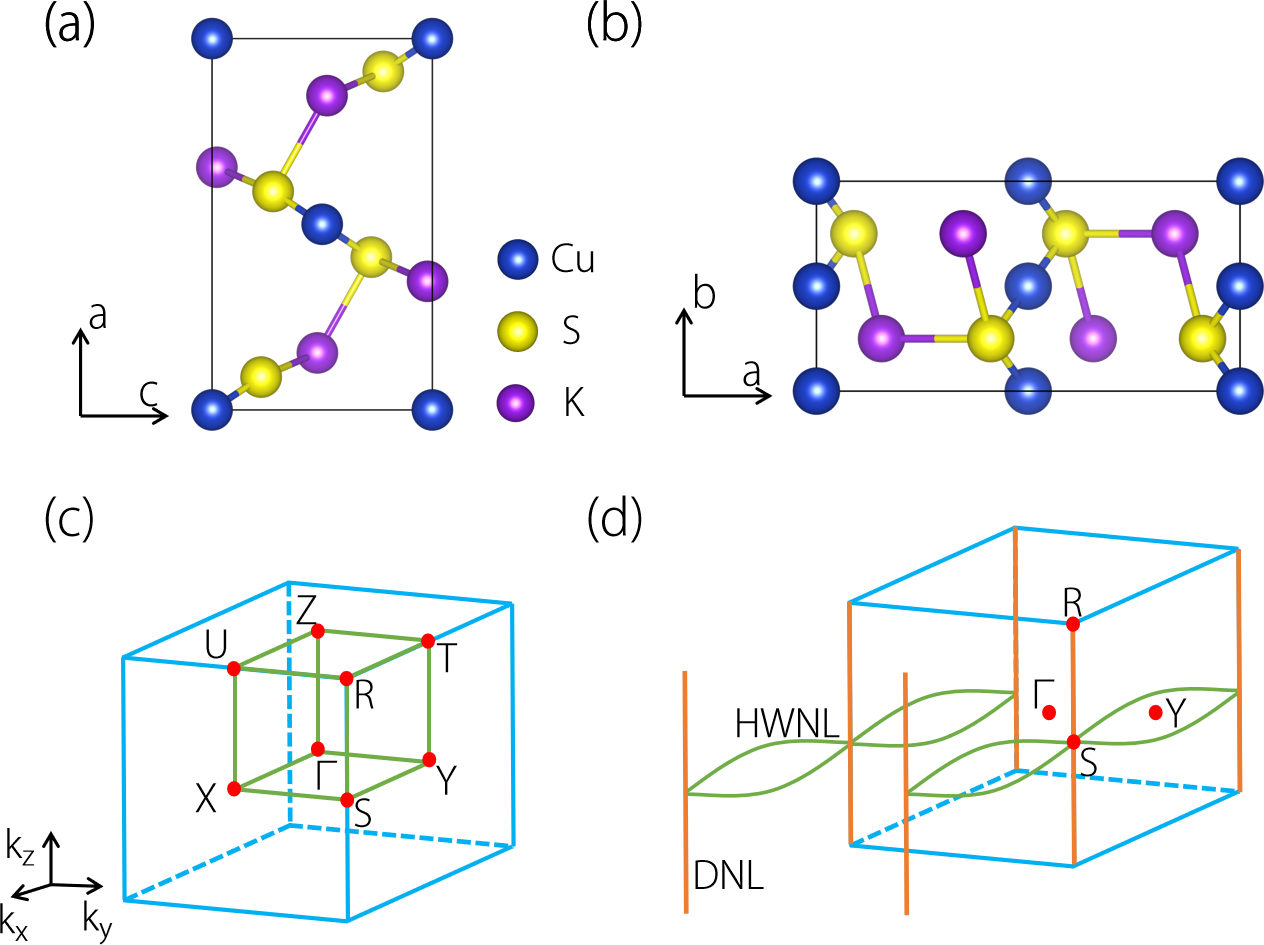}
\caption{(a) and (b) crystal structures of KCuS with $Pnma$-type structures under different viewsides; (c) 3D coordinate system of Brillouin zone (BZ) and some high symmetry points; (d) schematic diagram of two-fold degenerate hourglass Weyl nodal line (HWNL) and four-fold degenerate Dirac nodal line (DNL) phonons in 3D BZ.
\label{fig1}}
\end{figure}

Although phononic Weyl nodal line (WNL) states~\cite{add53,add54} have been proposed in some solid-state materials before, to the best of our knowledge, the phononic DNL states have rarely been predicted by other researchers. More importantly, a natural question is whether two-fold degenerate WNL phonons and four-fold degenerate DNL phonons can coexist in one single solid-state material. In this paper, we answer this question with certainty. For the first time, via first-principles calculations and symmetry analysis, we propose that KCuS with $Pnma$ symmetry group is a topological phononic material exhibiting Weyl and DNL phonons. Remarkably, the predicted nodal line phonons are special in that (i) the two-fold degenerate WNL states are formed by the neck crossing point of the hourglass-like dispersion~\cite{add56,add57,add58,add59,add60}; (ii) the four-fold degenerate DNLs belong to open nodal lines; (iii) the Dirac and Weyl phonon bands are nearly flat and the only ``clean'' bands in the 5.0--5.2-THz range; (iv) phonon surface states occur only in the [100] and [001] surfaces, unlike those of typical phononic nodal-line materials; and (v) KCuS provides a good platform to study the entanglement between the HWNL phonons and the DNL phonons.

\begin{figure}
\includegraphics[width=7.8cm]{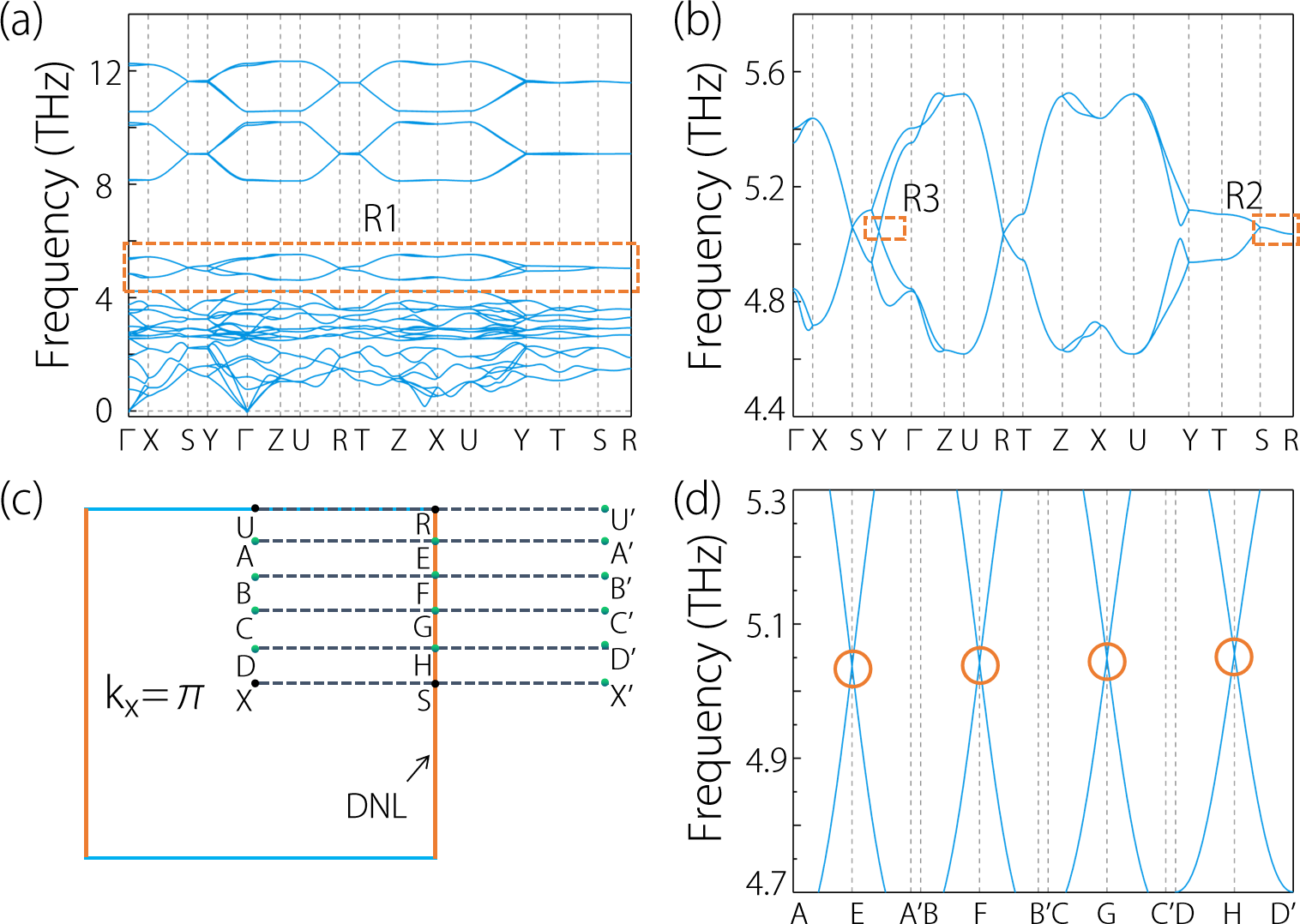}
\caption{ (a) phonon dispersion of $Pnma$ KCuS along $\Gamma$--X--S--Y--$\Gamma$--Z--U--R--T--Z--X--U--Y--T--S--R paths and (b) enlarged figure of R1 region in (a); (c) some selected symmetry points in $k_x=\pi$ plane; (d) calculated phonon dispersions along A--E--A$^{\prime}$, B--F--B$^{\prime}$, C--G--C$^{\prime}$, and D--H--D$^{\prime}$ paths, where the linear phonon band crossing points are marked with orange circles.
\label{fig2}}
\end{figure}

\section{Computational Methods}
We used density functional theory to calculate the ground state of KCuS with a $Pnma$ structure and the GGA-PBE ~\cite{add61} formalism for the exchange-correlation functional. We used the projector augmented-wave method for the interactions between ions and valence electrons and set the energy cutoff to 600 eV. We used a $\Gamma$--centered k-mesh of 5$\times$5$\times$5 for BZ sampling. We performed the lattice dynamic calculations to obtain the phonon dispersion of KCuS at its equilibrium lattice constants in the PHONOPY package ~\cite{add62} using density functional perturbation theory. We simulated the topological behaviors of the [100], [010], and [001] phonon surface states by constructing a Wannier tight-binding Hamiltonian of phonons ~\cite{add63}.

\section{Results and Discussion}
Figs.~\ref{fig1}(a) and~\ref{fig1}(b) show the crystal structure of KCuS with a $Pnma$-type structure under different viewsides. KCuS contains 12 atoms (i.e., four Cu, four K, and four S atoms located at the 4a (0.0, 0.5, 0.0), 4c (0.84639, 0.75, 0.52184), and 4c (0.41155, 0.25, 0.72286) Wyckoff positions, respectively). The lattice constants of KCuS are obtained with the help of first-principles calculations, and the values are a = 10.726  {\r{A}}, b = 5.3087  {\r{A}}, and c = 6.348  {\r{A}}, respectively. Due to the KCuS hosts orthorhombic-type crystal structure, there are nine independent elastic constants: $C_{11}$, $C_{12}$, $C_{13}$, $C_{22}$, $C_{23}$, $C_{33}$, $C_{44}$, $C_{55}$, $C_{66}$. These above-mentioned independent elastic constants are equal to 19.929 GPa, 13.589 GPa, 15.712 GPa, 33.184 GPa, 14.278 GPa, 18.756 GPa, 7.002 GPa, 7.883 GPa, and 3.081 GPa, respectively, according to our calculations. These elastic constants obey the following elastic stability criteria~\cite{add64}:
 \begin{equation}\label{1}
   \left\{
   \begin{aligned}
  &C_{11}>0;~C_{11} \times C_{22}>C_{12}^{2};\\
  &C_{11} \times C_{22} \times C_{33}+2C_{12} \times C_{13} \times C_{23}\\
  &~~-C_{11} \times C_{23}^{2}-C_{22} \times C_{13}^{2}-C_{33} \times C_{12}^{2}>0;\\
  &C_{44}>0;~C_{55}>0;~C_{66}>0
  \end{aligned}
  \right.
 \end{equation}
Hence, it can be concluded that $Pnma$ KCuS is mechanically stable.

Based on the selected high symmetry points in Fig. ~\ref{fig1}(c), we determined the dynamical stability of $Pnma$ KCuS through the phonon dispersion calculations. The phonon dispersion of KCuS along the $\Gamma$--X--S--Y--$\Gamma$--Z--U--R--T--Z--X--U--Y--T--S--R paths is shown in Fig.~\ref{fig2}(a). Obviously, the absence of imaginary frequency modes in the phonon dispersion indicates that KCuS is dynamically stable. We focused on the four phonon bands appearing in the 4.6-5.6 THz range (labeled R1 region); region R1 is well separated from the other phonon bands. For clarity, the enlarged phonon frequency of R1 is shown in Fig.~\ref{fig2}(b), and we divided R1 into two regions, namely, R2 and R3. In the R2 region, the phonon bands along the S-R path are four-fold degenerate; however, in the R3 region, one two-fold degenerate phonon band crossing point along the Y--$\Gamma$ path can be found. For $Pnma$-type KCuS, its symmetry operators are summarized as follows: two screw rotations $\widetilde{C_{2z}}=\{C_{2z}|\frac{1}{2}0\frac{1}{2}\}$ and $\widetilde{C_{2y}}=\{C_{2y}|0\frac{1}{2}0\},$ a spatial inversion \textit{P}, and time-reversal symmetry ${\cal{T}}$ with ${\cal{T}}^2=1$ (since it is a spinless system).

We first come to study the four-fold degenerate nodal line phonons along the S-R path. KCuS hosts three orthogonal two-fold screw rotation axes. Considering a combined antiunitary operation $\cal{T}$$\widetilde{C_{2i}}$,$(i=x,y,z),$ one can easily derive that $(\cal{T}$$\widetilde{C_{2\textit{i}}})^2=e^{\textit{ik}_{\textit{i}}}.$ Consequently, at the corresponding plane, $k_i=\pi,$ one has $(\cal{T}$$\widetilde{C_{2\textit{i}}})^2=-1.$ That is, the phonon dispersions along all boundary planes $(k_{x/y/z}=\pi)$ are at least two-fold degenerate. Therefore, the whole $k_{x/y/z}=\pi$ planes are covered by nodal surface phonons~\cite{add55}. As shown in Fig.~\ref{fig2}(b), one can see that there are two double degenerate nodal lines along Y-T-S, and these two nodal lines are formed into one four-fold degenerate nodal line along the S-R path. Hence, the four-fold degenerate DNL (e.g., S-R path) is sitting at the hinge between the $k_x=\pi$ and $k_y=\pi$ planes. That is, the phononic DNL (e.g., S-R path) is formed by the crossing of the phononic nodal surface states of $k_x=\pi$ and $k_y=\pi$ planes. Note that the DNL phonons along S-R belong to open nodal line states~\cite{add54,add55,add56,add57,add58,add59,add60,add61,add62,add63,add64,add65}, as shown in Fig.~\ref{fig2}(c). More interestingly, as shown in Fig.~\ref{fig2}(c) and Fig.~\ref{fig2}(d), we selected a series of phonon band crossing points (E, F, G, and H points) along the S-R path and showed the phonon dispersions along the A--E--A$^{\prime}$, B--F--B$^{\prime}$, C--G--C$^{\prime}$, and D--H--D$^{\prime}$ paths. Obviously, it can be concluded that the DNL along the S-R path is formed by a series of linear phonon band crossing points. These phonon band crossing points are nearly flat with small frequency variation (see the orange circles in Fig.~\ref{fig2}(d); the orange circles are almost flat in frequency), and exhibit a large linear frequency range.

To further prove the occurrence of the phononic DNL along the S-R path, the symmetry analysis is shown as follows: The DNL lies at the hinge between two planes (e.g., $k_x=\pi$ and $k_y=\pi$); it is the invariant subspaces of $\widetilde{C_{2z}}$, $\widetilde{M_y}$, and a combined operation $\cal{T}$$\widetilde{C_{2y}}$. The commutations between them are given by,\begin{equation}\label{2}
\widetilde{C_{2z}}\widetilde{M_y}=\mathcal{T} _{010}\widetilde{M_y}\widetilde{C_{2z}},
\end{equation}
\begin{equation}\label{3}
\widetilde{M_y}(\mathcal{T} \widetilde{C_{2y}})=(\mathcal{T} \widetilde{C_{2y}})\widetilde{M_{y}},
\end{equation}where $\mathcal{T} _{010}$ is the translation along the y-direction.
\\Along the S-R path, one has \begin{equation}\label{4}
\{\widetilde{C_{2z}},\widetilde{M_y}\}=0, [\widetilde{M_y},(\mathcal{T} \widetilde{C_{2y}})]=0.
\end{equation}The Bloch states along this path can be chosen as the eigenstates of $\widetilde{M_y}$, characterized by its eigenvalues, $|g_y=\pm1\rangle$. Due to Eq.(\ref{3}), $|g_y=1\rangle$ and $\mathcal{T} \widetilde{C_{2y}}|g_y=1\rangle$ are degenerate due to a Kramer-like degeneracy, with $(\mathcal{T} \widetilde{C_{2y}})^2=-1$ at this path. In addition, according to Eq.(\ref{2}), the anticommutation relationship also implies another degeneracy; $|g_y=1\rangle$ and $\widetilde{C_{2z}}|g_y\rangle$ are degenerate, corresponding to the opposite $g_y$. Consequently, $\{|g_y=1\rangle,|g_y=-1\rangle, \mathcal{T} \widetilde{C_{2y}}|g_y=1\rangle,T\widetilde{C_{2y}}|g_y=-1\rangle\}$ are degenerate along this path; there is indeed a four-fold degenerate line (i.e., a DNL) along the S-R path.

\begin{figure}
\includegraphics[width=7.8cm]{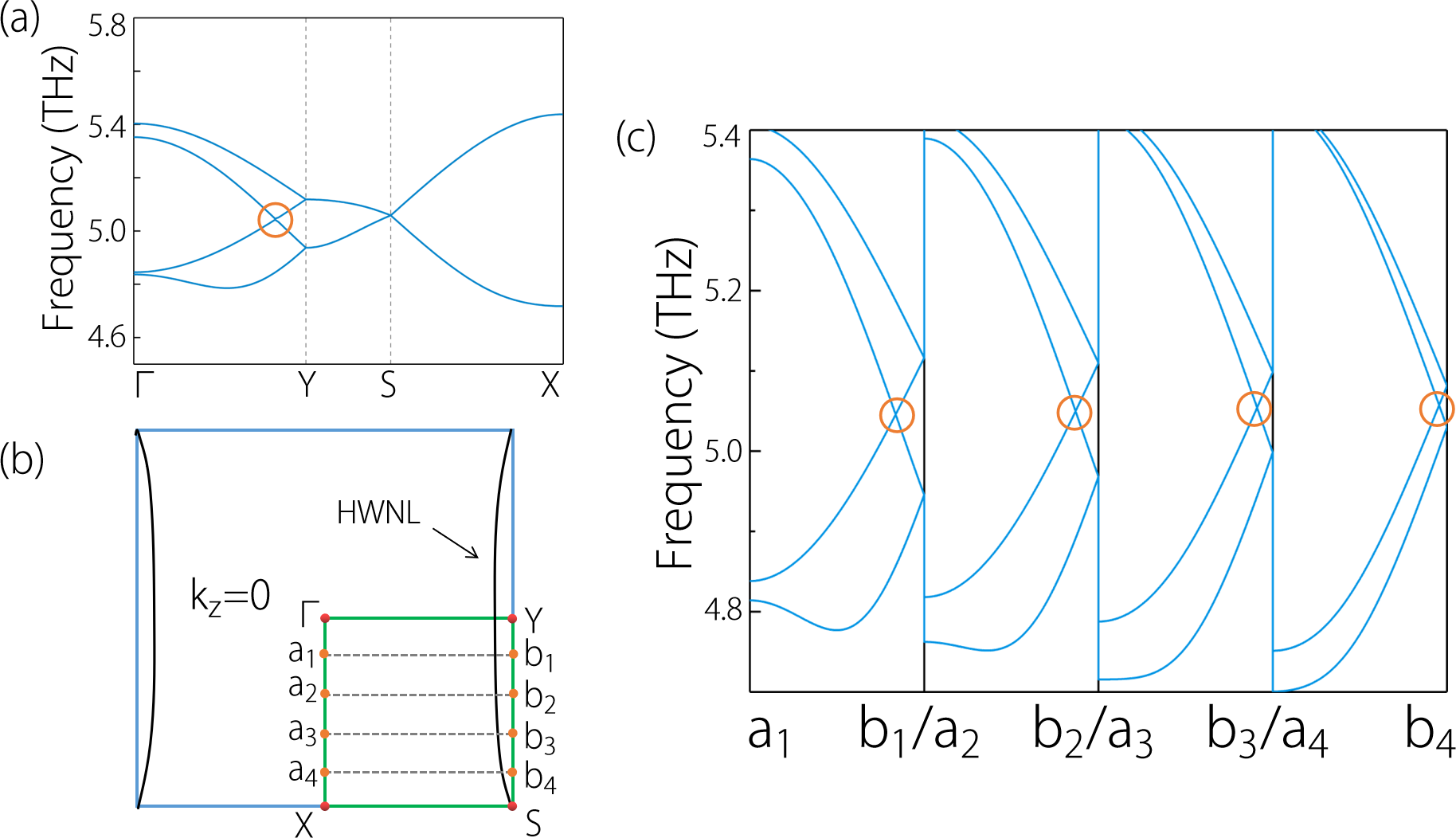}
\caption{(a) enlarged phonon dispersion of $Pnma$ KCuS in R3 region of Fig.~\ref{fig2}(b) where hourglass-type phonon band crossing along $\Gamma$-Y path is marked with circle; (b) series of selected symmetry points in $k_z=0$ plane; (c) detailed phonon dispersions along a$_m$-b$_m$ (m = 1, 2, 3, 4) paths where hourglass-type phonon band crossings are marked with orange circles.
\label{fig3}}
\end{figure}

We now come to study the two-fold degenerate phonon band crossing point (see Fig.~\ref{fig3}(a)) along the Y-$\Gamma$ path in R3. $k_z=0$ plane is an invariant subspace of $\widetilde{M_z}$, and the Bloch states on it can be characterized by its eigenvalues $g_z=\pm e^{{ik_x}/{2}}$. According to the Kramer-like degeneracy, we know that the $k_i=\pi$ $(i=x,y,z)$ plane has double degeneracy. At the Y point, which lies on plane $k_y=\pi$, one has $[\widetilde{M_z},(\mathcal{T} \widetilde{C_{2y}})]=0$ and $g_z=\pm1$, so that $\{|g_z=1\rangle,|g_z=1\rangle\}$ are degenerate at this point. In contrast, at the P point (see Fig. S1), a generic point along X-S, lying on the $k_x=\pi$ plane, one has $[\widetilde{M_z},(\mathcal{T} \widetilde{C_{2x}})]=0$, but $g_z=\pm i$, such that $\{|g_z=+i\rangle,|g_z=-i\rangle\}$ are degenerate. Consequently, from P to Y, there must be a doublet switching (see in Fig. S1), leading to a neck crossing point of the hourglass-like dispersion. Due to the presence of $\widetilde{M_z}$, such a crossing point is not isolated but inducing a line within the $k_z=0$ plane, as shown in Fig.~\ref{fig3}(b). The shape of the nodal line in the $k_z=0$ plane is shown in Fig.~\ref{fig3}(b) using black lines. Moreover, we selected a series of symmetry points, a$_1$-a$_4$ (b$_1$-b$_4$), between the $\rm{\Gamma(Y)}$ and X (S) points to show the detailed phonon dispersions along the a$_m$-b$_m$ (m = 1, 2, 3, 4) paths. The results are shown in Fig.~\ref{fig3}(c); along the a$_m$-b$_m$ paths, a series of phonon band crossing points appear, and they belong to the neck crossing points of the hourglass-like dispersion~\cite{add56,add57,add58,add59,add60}. Therefore, one can see that the two-fold degenerate nodal line phonons in the $k_z=0$ plane should be HWNL phonons. Normally, the hourglass nodal line (HNL) can move in the $k_z=0$ plane; however, the HWNL as exhibited in Fig.~\ref{fig3}(b) is unmovable, as its two endpoints are pinned at the S point. Moreover, as shown in Fig.~\ref{fig3}(c), one can see that the hourglass-type phonon band crossings (see orange circles) are almost flat in frequency.

\begin{figure}
\includegraphics[width=7.8cm]{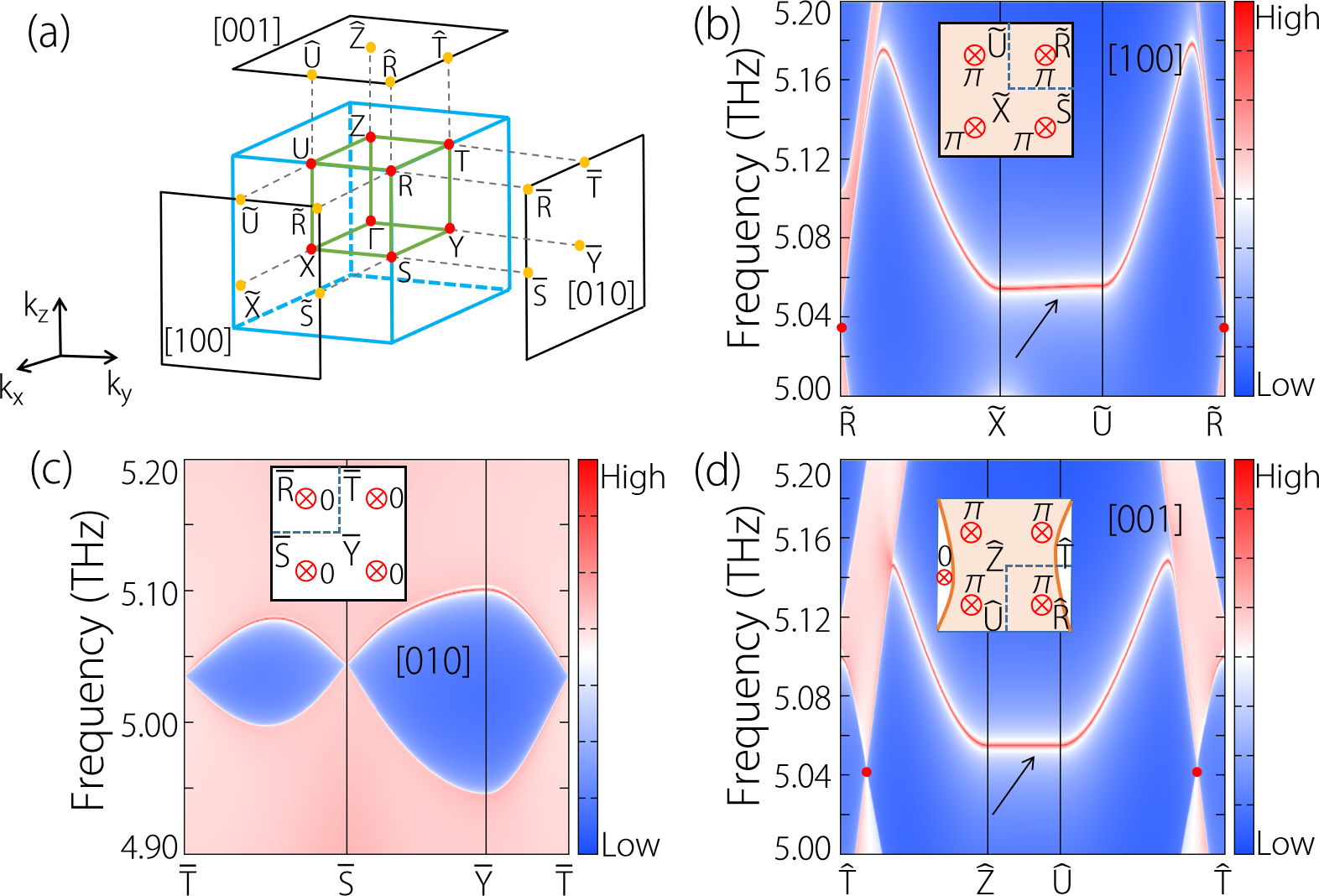}
\caption{(a) 3D bulk BZ and [001], [010], [100] surface BZs; (b)-(d) phonon surface spectra on [100], [010], [001] surfaces, respectively. The inset figures in (b)-(d) indicate the values of Zak phases (0 or $\pi$) for lines normal to the [100], [010], [001] surfaces. The nontrivial phonon surface states in the [100] and [001] surfaces are indicated by black arrows. For the [010] surface, no phonon surface states appear.
\label{fig4}}
\end{figure}

Finally, we start to study the phonon surface spectra of $Pnma$-type KCuS. As shown in Fig.~\ref{fig4}, one can conclude that the phonon surface states can only appear in the [100] and [001] surfaces. However, for the [010] surface, no phonon surface state can be observed (see Fig.~\ref{fig4}(c)). To understand the physics, we employed Zak phase calculations in this work. The Zak phase is the Berry phase of a straight line normal to the surface and crossing the bulk BZ. A $\pi$ Zak phase generally indicates the existence of nontrivial topological surface states. For lines normal to the [100] and [001] surfaces, the Zak phase equals $\pi$ (see the inset figures of Figs.~\ref{fig4}(b) and~\ref{fig4}(d)), corresponding to the appearance of surface states. However, for a line normal to the [010] surface, the Zak phase equals 0 (see the inset figure of Fig.~\ref{fig4}(c)), corresponding to the disappearance of surface states.

\section{Conclusions}
In summary, based on first-principles calculations and symmetry analysis, nearly flat phononic DNL and phononic HWNL states were predicted in one solid-state material, KCuS with a $Pnma$ structure. The $Pnma$-type KCuS was predicted to be mechanically and dynamically stable in terms of theory. Moreover, the phonon surface states in the [100], [010], and [001] surfaces of this material were investigated; according to the results, the topological phonon surface states of KCuS only appeared in the [100] and [001] surfaces due to the $\pi$ Zak phases. The current work predicted the appearance of coexisting two-fold and four-fold degenerate nodal line phonons in a single material for the first time. The KCuS reported here therefore can be viewed as a good platform to study the entanglement between DNL phonons and HWNL phonons in the future.

\emph{\textcolor{blue}{Acknowledgments}} X.T. Wang is grateful for the support from the National Natural Science Foundation of China (No. 51801163) and the Natural Science Foundation of Chongqing (No. cstc2018jcyjA0765).

\end{document}